# Noncontact atomic force microscopy: Stability criterion and dynamical responses of the shift of frequency and damping signal

G. Couturier,[a)] R. Boisgard, L. Nony,[b)] and J. P. Aimé
*Centre de Physique Moléculaire Optique et Hertzienne, Université Bordeaux I, UMR5798 CNRS, 351 Cours de la Libération, 33405 Talence Cedex, France*

The aim of this article is to provide a complete analysis of the behavior of a noncontact atomic force microscope (NC-AFM). We start with a review of the equations of motion of a tip interacting with a surface in which the stability conditions are first revisited for tapping mode. Adding the equations of automatic gain control (AGC), which insures constant amplitude of the oscillations in the NC-AFM, to the equations of motion of the tip, a new analytical stability criterion that involves proportional and integral gains of AGC is deduced. Stationary solutions for the shift of frequency and for the damping signal are obtained. Special attention is paid to the damping signal in order to clarify its physical origin. The theoretical results are then compared to those given by a virtual machine. The virtual machine is a set of equations solved numerically without any approximation. The virtual machine is of great help in understanding the dynamical behavior of the NC-AFM as images are recorded. Transient responses of the shift in frequency and of the damping signal are discussed in relation to the values of proportional and integral gains of AGC.

## I. INTRODUCTION

The noncontact atomic force microscope (NC-AFM) is a powerful tool with which to investigate surface properties at the nanometer scale. Contrast at the atomic scale has been achieved for semiconductors and insulators.[1–6] The NC-AFM is also a powerful tool with which to investigate soft materials.[7] Albretch *et al.* were the first to propose the concept of a NC-AFM.[8] In the NC-AFM, the tip–cantilever is in a closed loop and the frequency of the oscillations depends on tip–surface interaction. The amplitude of the oscillations is kept constant by automatic gain control (AGC). The damping signal, which is the signal of error of AGC, should normally be a good measure of the dissipative term involved in tip–surface interaction. The question of the physical origin of the variation of the damping signal remains unresolved and has been a matter of debate over the last few years.[9–13] This point is of great importance because the damping signal should provide information on dynamical properties at the molecular scale. With the development of dynamic force microscopy (DFM), in tapping or noncontact mode, numerous theoretical work[14–19] has been devoted to a description of the dynamical behavior of an oscillating tip–cantilever system (OTCS) in proximity to a surface. Analytical solutions and numerical results predict and show stable and unstable domains of the resonance peak in tapping mode, i.e., when the OTCS is excited by a constant driving force. Theoretical predictions are widely confirmed by experimental results.[20] However, it is not clear whether the NC-AFM can be understood by extrapolating theoretical results obtained by assuming a constant driving force like done in tapping mode. Unlike tapping mode, NC-AFM mode looks very stable: the phase of the OTCS may be adjusted at any value around $-\pi/2$ whereas theoretical results predict instability for phase larger than $-\pi/2$ in tapping mode.

To understand a NC-AFM machine, we need to add the equations that rule AGC to the equations of motion of the OTCS. This article is organized as follows. In Sec. II, the equations of motion of the OTCS without AGC and the stability criterion are reviewed. In Sec. III, the equations that rule AGC are added to the equations of motion of the OTCS in order to establish the domain of stability. We show that noncontact mode is stable as long as proportional and integral gains of AGC satisfy three inequalities. The stationary solutions for the shift of frequency and for the damping signal versus the tip–surface distance are also established for any value of the phase of the OTCS. Theoretical results are obtained with some minor approximations, and the predictions are compared to the results given by the virtual NC-AFM machine. The virtual machine is made of a set of nonlinear differential equations solved numerically without any approximation; this machine is a model of the machine used in our laboratory and also in most other laboratories. The virtual machine can calculate the transient response of the shift in frequency and of the damping signal when nondissipative or/and dissipative force is introduced into tip–surface interaction. Transient responses are strongly related to proportional and integral gains of AGC and may lead, in some

[a)] ~~Author to whom correspondence~~ should be addressed; electronic mail: gcoutur@cribx1.u-bordeaux.fr
[b)] Current address: Zurich Research Labatory, IBM Research, 8803 Rüschlikon, Switzerland.

cases, to misinterpretation of the NC-AFM images. Typical artifacts are thus discussed at the end of the article.

## II. BEHAVIOR OF AN OSCILLATING TIP EXCITED BY CONSTANT DRIVING FORCE: TAPPING MODE

Here in Sec. II, we briefly recall the main results for an oscillating tip close to a surface. In so-called tapping mode, oscillations of the tip are driven by an external force at a given frequency.

The differential equation that describes motion $z(t)$ of the tip is given by

$$m^*\frac{d^2z(t)}{dt^2}+\frac{m^*\omega_0}{Q}\frac{dz(t)}{dt}+k_cz(t)=F_{\rm ext}(t)-\nabla V_{\rm int}[z(t)], \tag{1}$$

where $\omega_0$, $m^*$, and $k_c=m\omega_0^2$ are, respectively, the resonant frequency, the effective mass, and the cantilever stiffness of the OTCS. $Q=1/2\gamma$ is the quality factor and $\gamma$ is the damping coefficient. $F_{\rm ext}(t)=F_0\cos(\omega t)$ is the external driving force, and $V_{\rm int}[z(t)]$ is the interaction potential between the tip and the surface. In this article, we assume (i) that the tip never touches the surface and (ii) van der Waals sphere–plane interaction,[21] thus $V_{\rm int}[z(t)]=-\{HR/6[D-z(t)]\}$ where $H$, $R$, and $D$ are the Hamaker constant, the tip's apex radius, and the tip–surface distance, respectively. This particular potential does not restrict the validity of the results discussed in this article.

To solve the nonlinear differential, Eq. (1), we used the principle of least action, so we start by building the Lagrangian $L(z,\dot z,t)=T-U+W$:

$$L(z,\dot z,t)=\frac{1}{2}m^*\dot z(t)^2-\left[\frac{1}{2}k_cz(t)^2-z(t)F_0\cos(\omega t)+V_{\rm int}[z(t)]\right]-\frac{m^*\omega_0}{Q}z(t)\underline{\dot z(t)}, \tag{2}$$

where the underlined variable $\dot z(t)$ is calculated along the physical path, and thus is not varied in the calculations.[22]

Due to the large value of the quality factor $Q$, we assume a typical temporal solution of the form

$$z(t)=Z_m(t)\cos[\omega t+\theta(t)], \tag{3}$$

where $Z_m(t)$ and $\theta(t)$ are assumed to be functions that slowly vary over time compared to the period $T=2\pi/\omega$.

Applying the principle of least action $\delta S=0$, with $S=\Sigma_n(1/T\int_{nT}^{(n+1)T}L(z,\dot z,t)dt)T$, we obtain the Euler–Lagrange equations. Thus, the amplitude and phase equations of motion of the OTCS coupled to the surface are obtained:[19]

$$\begin{aligned}\frac{\ddot Z_m}{\omega_0^2}&=\left[\left(u+\frac{\dot\theta}{\omega_0}\right)^2-1\right]Z_m-\frac{\dot Z_m}{\omega_0 Q}+\frac{F_0}{k_c}\cos(\theta)+\frac{Z_m\kappa}{3(D^2-Z_m^2)^{3/2}},\\ \frac{\ddot\theta}{\omega_0^2}&=-\left(\frac{2\dot Z_m}{\omega_0 Z_m}+\frac{1}{Q}\right)\left(u+\frac{\dot\theta}{\omega_0}\right)-\frac{F_0}{k_c}\frac{\sin(\theta)}{Z_m},\end{aligned} \tag{4}$$

where $u=\omega/\omega_0$ and $\kappa=HR/k_c$.

The equations of motion of the stationary solutions $Z_{ms}$ and $\theta_s$ are obtained by setting $\ddot Z_m=\ddot\theta=\dot Z_m=\dot\theta=0$ in Eq. (4):

$$\begin{aligned}Z_{ms}(1-u^2)-\frac{Z_{ms}\kappa}{3(D^2-Z_{ms}^2)^{3/2}}&=\frac{F_0}{k_c}\cos(\theta_s),\\ Z_{ms}\frac{u}{Q}&=-\frac{F_0}{k_c}\sin(\theta_s).\end{aligned} \tag{5}$$

From Eqs. (5), we derived the equations that give the shape of the resonance peak and the phase as a function of the distance $D$ and the force $F_0$:

$$\omega_\pm=\omega_0\sqrt{\left(\frac{QF_0}{k_c}\right)^2\frac{1}{Z_{ms}^2}-\frac{1}{4Q^2}\left\{1\mp\sqrt{1-4Q^2\left[1-\left(\frac{QF_0}{k_c}\right)^2\frac{1}{Z_{ms}^2}-\frac{\kappa}{3(D^2-Z_{ms}^2)^{3/2}}\right]}\right\}^2},$$

$$\theta_{s\pm}=\arctan\left(\frac{u_\pm}{Q(u^2-1)+\dfrac{Q\kappa}{3(D^2-Z_{ms}^2)^{3/2}}}\right). \tag{6}$$

Typical plots of $Z_{ms}$ and $\theta_s$ versus the frequency $f=\omega/2\pi$ are shown in Figs. 1(a) and 1(b). As already discussed in various papers,[18,19] the nonlinearity of the interaction greatly distorts the resonance peak, and two branches appear. Equation (1) is similar to the Duffing equation that has been studied extensively.[23–25] It is well known that, by sweeping the frequency, the amplitude and phase exhibit jumps at frequency where the derivative of $Z_{ms}$ diverges. From Eq. (1), the stability is deduced by substituting $Z_m$ and $\theta$ in Eqs. (4) by $(Z_{ms}+m)$ and $(\theta_s+p)$, where $m$ and $p$ are infinitesimal. A second order differential equation of the variable $m$ is then obtained. Using, for instance, the Rooth–Hurwitz stability criterion,[26] the stability is given by

$$g(Z_{ms},u,Q,\kappa,D,F_0,k_c)>0,$$

where

$$g=\frac{Z_{ms}u^2}{Q^2}+\left[Z_{ms}(1-u^2)-\left(\frac{Z_{ms}\kappa}{3(D^2-Z_{ms}^2)^{3/2}}\right)\right]\times\left[(1-u^2)-\frac{QF_0}{k_c}\frac{d}{dZ_m}\left(\frac{Z_m\kappa}{3(D^2-Z_m^2)^{3/2}}\right)_{Z_m=Z_{ms}}\right]. \tag{7}$$

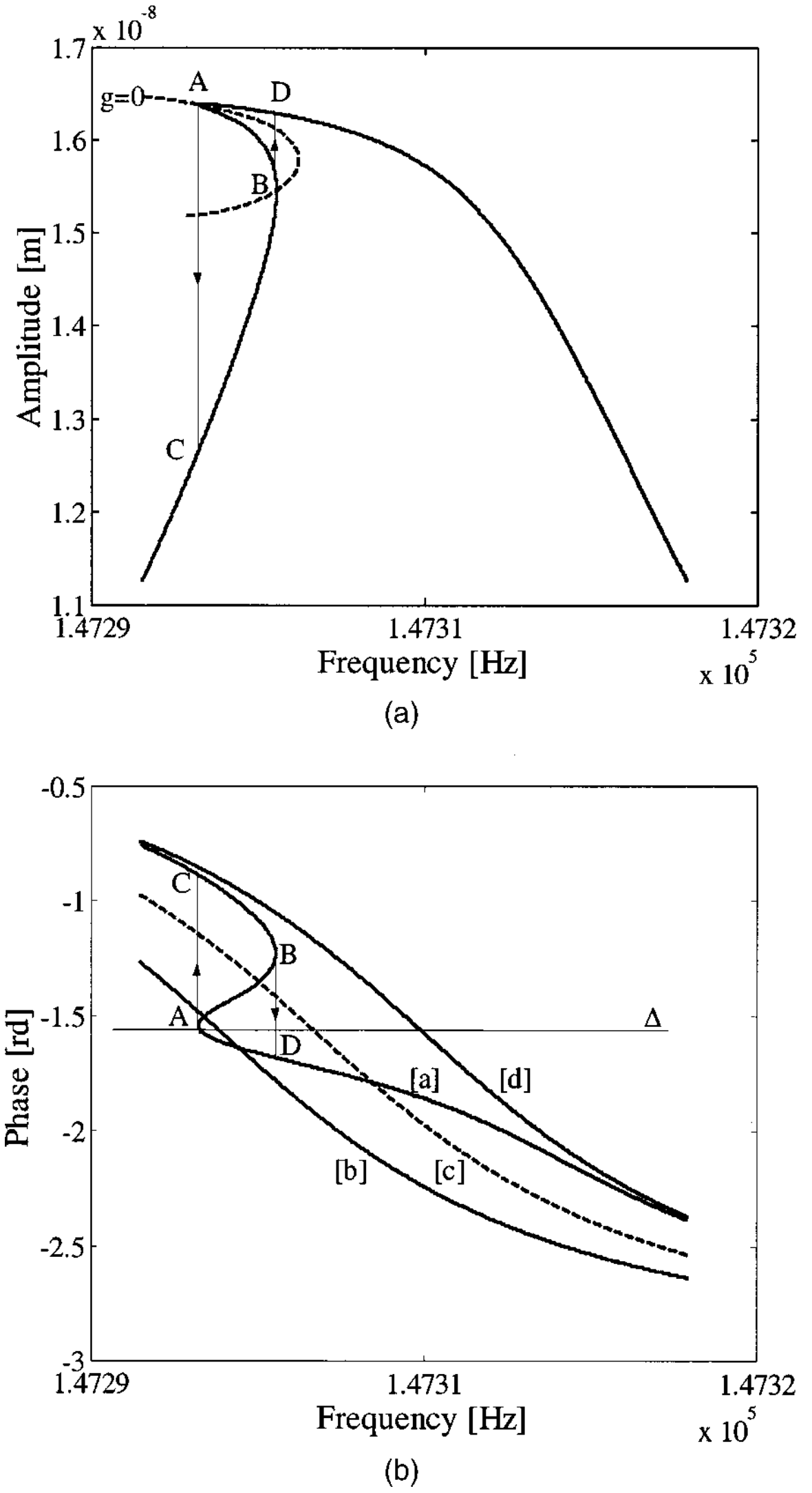

FIG. 1. Distortion of the resonance peak (a) and phase [curve a in (b)] vs the frequency for an OTCS without CAG ($F_0 = 1.38\times10^{-10}$ N, $Q = 4750$, $D = 17$ nm, $f_0 = 147\,305$ Hz, and $\kappa = 5\times10^{-29}$ m$^3$). Curves b and c are the phase curves [Eq. (14)] of the OTCS with CAG for $Z_{ms} = 16.35$ and 16 nm, respectively, and $D = 17$ nm; curve d is the phase for $D\to\infty$, it does not depend on $Z_{ms}$.

We have to point out that the terms $\ddot{Z}_m/\omega_0^2$, $\dot{Z}_m/\omega_0 Q$, $Z_m\dot{\theta}^2/\omega_0^2$, and $\ddot{\theta}/\omega_0^2$, $2\dot{Z}_m\dot{\theta}/Z_m\omega_0^2$, $\dot{\theta}/Q\omega_0$ were neglected in Eqs. (4); this approximation is supported by the fact that (i) $Z_m(t)$ and $\theta(t)$ are assumed to be slowly varying functions over time and (ii) the quality factor $Q$ is high. Numerical resolution of Eqs. (4) shows that the approximation is completely reasonable.

The plot of $g(Z_{ms},u,Q,\kappa,D,F_0,k_c) = 0$ is shown by dashed lines in Fig. 1(a) and plot $g = 0$ crosses plot $Z_{ms}$ versus frequency at points $A$ and $B$ where $[dZ_{ms}/d\omega]^{-1} = 0$. Thus, when the frequency is swept from high to low values, the amplitude $Z_{ms}$ and phase $\theta_s$ jump from $A$ to $C$; conversely, sweeping from low to high values gives jumps of $Z_{ms}$ and $\theta_s$ from $B$ to $D$. These jumps have been already observed in tapping mode by various authors.[15,20,27]

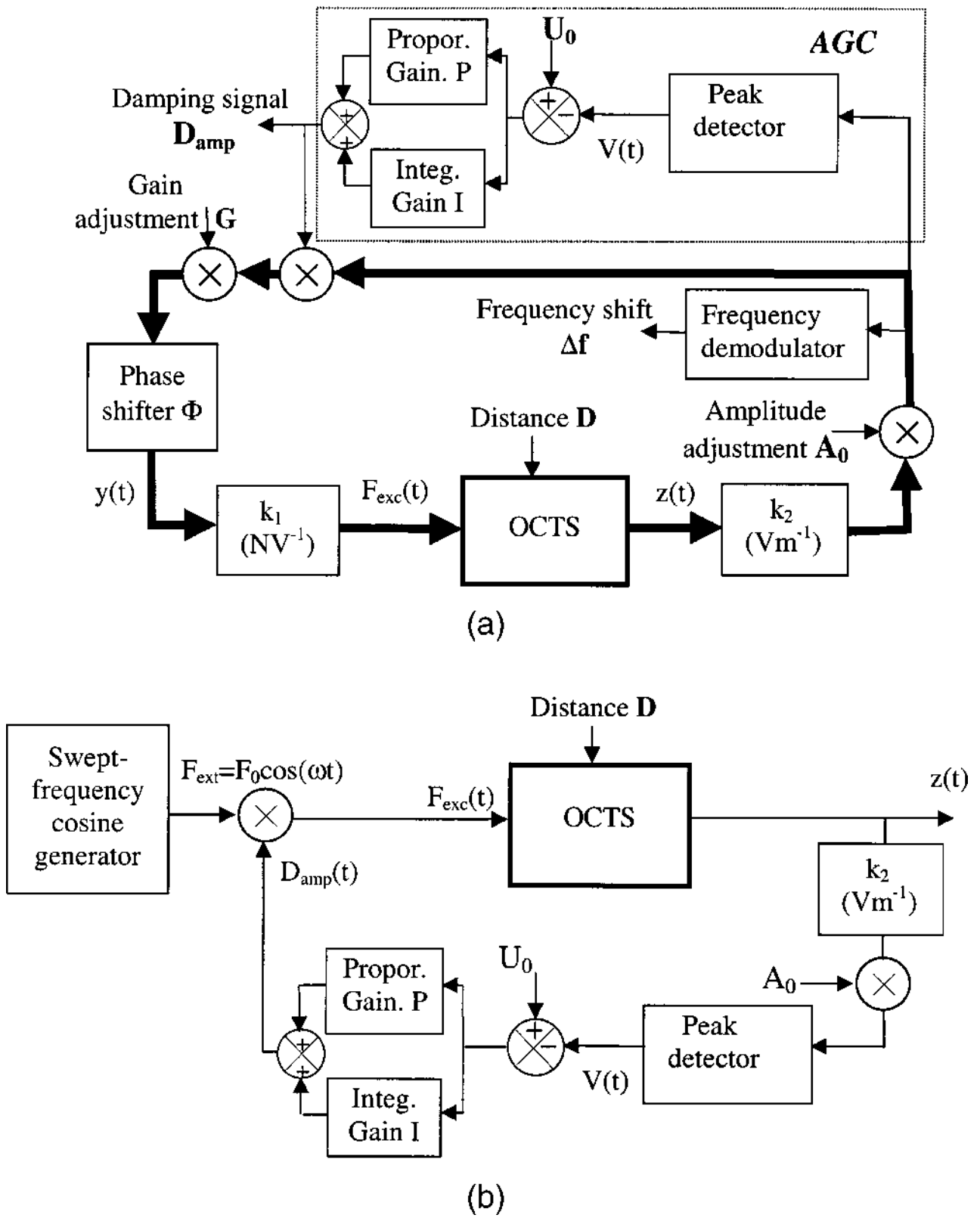

FIG. 2. Schematic diagram of a NC-AFM (a). (b) Schematic diagram of the OTCS with CAG used to establish the domain of stability in the NC-AFM.

## III. BEHAVIOR OF AN OSCILLATING TIP KEPT AT CONSTANT AMPLITUDE: NONCONTACT MODE

### A. Equations of motion

First, we will briefly recall the principle of the NC-AFM. A simplified schematic diagram of the microscope is given in Fig. 2(a). The OTCS is in a closed loop and the amplitude of the oscillations is adjusted by $A_0$ and is kept at a constant level by AGC. The so-called damping signal $D_{\mathrm{amp}}$ is the signal of error of AGC. $k_1$ (in N V$^{-1}$) and $k_2$ (in V m$^{-1}$) are the apparatus' functions of cantilever piezo and of optical detection, respectively. The frequency of oscillations of the loop is measured by a quadrature frequency demodulator. The phase shift network ($\phi$) and the gain $G$ are adjusted to satisfy the Barkhausen criterion[28] at the oscillation frequency selected, in other words, the gain in open loop is equal to unity at the frequency of the oscillations. Thus, one way to understand the stability in noncontact mode is to study the OCTS feedback through AGC as shown in Fig. 2(b). The equations that rule the OCTS are identical to Eqs. (4), except that the excitation $F_0$ is now replaced by the product $F_0 D_{\mathrm{amp}}$:

$$\frac{\ddot{Z}_m}{\omega_0^2} = \left[\left(u + \frac{\dot{\theta}}{\omega_0}\right)^2 - 1\right] Z_m - \frac{\dot{Z}_m}{\omega_0 Q} + \frac{F_0 D_{\mathrm{amp}}}{k_c}\cos(\theta) + \frac{Z_m \kappa}{3(D^2 - Z_m^2)^{3/2}},$$

$$\frac{\ddot{\theta}}{\omega_0^2} = -\left(\frac{2\dot{Z}_m}{\omega_0 Z_m} + \frac{1}{Q}\right)\left(u + \frac{\dot{\theta}}{\omega_0}\right) - \frac{F_0 D_{\mathrm{amp}}}{k_c}\frac{\sin(\theta)}{Z_m}. \qquad (8)$$

The damping signal $D_{\rm amp}$ is given by

$$D_{\rm amp}=P(U_0-V)+\int_0^t I[U_0-V(u)]du \tag{9}$$

and its derivative by

$$\dot{D}_{\rm amp}=-P\dot{V}+I(U_0-V), \tag{10}$$

where $P$ and $I$ are proportional and integral gains, respectively, and $U_0$ is a constant. From a practical point of view, the peak detector is realized by means of a rectifying diode, a $R-C$ circuit, and a first-order low pass filter or a quadratic detector and a second-order low pass filter, so $V(t)$ is approximately ruled by a second-order differential equation:

$$\frac{1}{\omega_c^2}\ddot{V}+\frac{1}{Q_c\omega_c}\dot{V}+V=k_2A_0Z_m, \tag{11}$$

where $f_c=\omega_c/2\pi$ is the center frequency and $Q_c$ is the quality factor ($Q_c=\sqrt{2}/2$ for a Butterworth filter, 0.577 for a Bessel, etc.).

The equations of motion of the stationary solutions $Z_{ms}$ and $\theta_s$ are obtained by setting $\ddot{Z}_m=\ddot{\theta}=\dot{Z}_m=\dot{\theta}=\dot{D}_{\rm amp}=\ddot{V}=\dot{V}=0$ in Eqs. (8)–(11):

$$Z_{ms}(1-u^2)-\frac{Z_{ms}\kappa}{3(D^2-Z_{ms}^2)^{3/2}}=\frac{F_0D_{\rm amp}}{k_c}\cos(\theta_s),$$
$$Z_{ms}\frac{u}{Q}=-\frac{F_0D_{\rm amp}}{k_c}\sin(\theta_s) \tag{12}$$

with

$$Z_{ms}=\frac{U_0}{k_2A_0} \tag{13}$$

and

$$\theta_s=-\mathrm{arcot}\left[\frac{(1-u^2)-[\kappa/3(D^2-Z_{ms}^2)^{3/2}]}{\frac{u}{Q}}\right]. \tag{14}$$

From Eq. (13), it clearly appears that the amplitude of the oscillations is adjusted by $A_0$.

Curves b and c in Fig. 1(b) are plots of $\theta_s$ versus frequency for two different values of set point $Z_{ms}$; in both, the phase varies continually and no jump is observed. Curve d corresponds to the case where the distance $D$ is infinite. As shown below and in Sec. IV, curves b, c, and d are helpful in understanding the behavior of the NC-AFM.

The values of the $P$ and $I$ gains that ensure the stability of the system in Fig. 2(b) are obtained by substituting $Z_m$, $\theta$, $D_{\rm amp}$, and $V$ in Eqs. (8), (10), and (11) for $(Z_{ms}+m)$, $(\theta_s+p)$, $(D_{\rm amp}+q)$, and $(V_s+v)$ where $m$, $p$, $q$, and $v$ are infinitesimal. A fifth-order differential equation of the variable $m$ is then obtained. Using the Rooth–Hurwitz stability criterion, the stability requires the following three conditions:

$$P<K_1,$$
$$I<P^2K_2+PK_3+K_4, \tag{15}$$
$$I^2K_5+I(P^2K_6+PK_7+K_8)+P^3K_9+P^2K_{10}+PK_{11}>0,$$

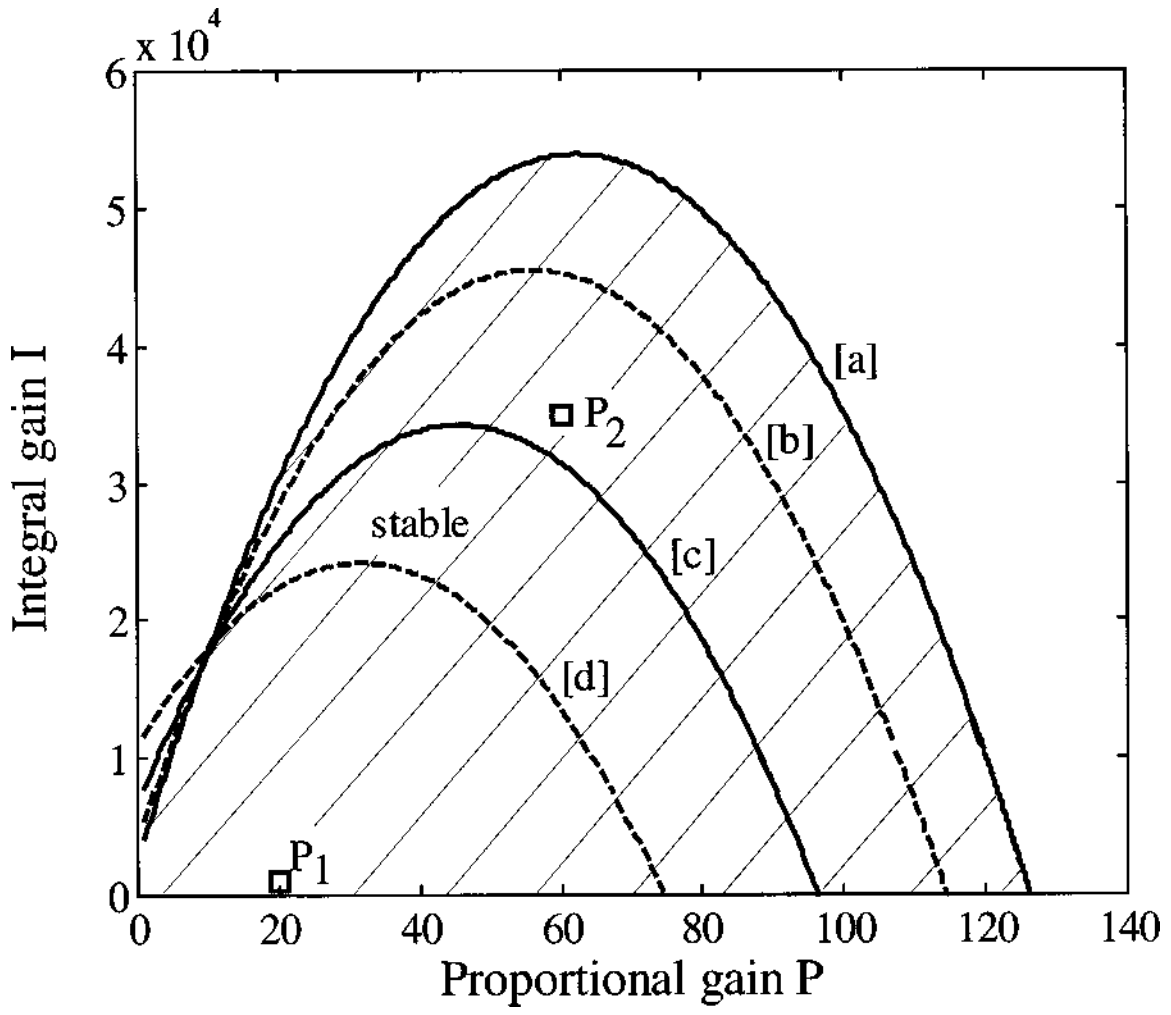

FIG. 3. Stability domain, i.e., integral gain $I$ vs proportional gain $P$ of CAG, for four different values of the distance $D$. Curves a, b, c, and d are for $D\to\infty$ and 15.18, 15.167, and 15.16 nm, respectively. The parameters for the calculations are $f_0=147\,305$ Hz, $Q=4750$, $k_c=40$ N m$^{-1}$, $U_0=0.638$, $Z_{ms}=15.1$ nm, $\kappa=5\times10^{-29}$ m$^3$, $\omega_c=3450$ rad s$^{-1}$, $Q_c=0.45$, and $F_{\rm ext}=1.270\times10^{-10}$ N.tability and dynamical response of AFM

where $K_1$, $K_2$,..., $K_{11}$ are functions of parameters $u$, $\omega_0$, $k_c$, $Q$, $U_0$, $k_2A_0$, $\omega_c$, $Q_c$, $D$, $F_0$, and $\kappa$.

Curve a and the $x$ axis in Fig. 3 delineate the stability domain when distance $D$ is infinite and $u=1$. The system would be stable if the $P$ and $I$ gains are chosen inside the hatched zone in Fig. 3. When adiabatic conditions are fulfilled, the stationary solutions of the shift in frequency and of the damping signal do not depend on the pair $(P-I)$; see, for instance, Eqs. (17) and (19) below. When the adiabatic conditions are not satisfied, the choice of pair $(P-I)$ is of great importance because transient responses of the shift in frequency and of the damping signal depend on the gain in $P$ and $I$. Unfortunately, Eqs. (8), (10), and (11) indicate that it is not easy to obtain a criterion that could help the user choose the gain in $P$ and $I$. The virtual machine becomes of some help in investigating this point as we will see in Sec. IV.

As shown in Fig. 3, the stability domain becomes smaller and smaller when the tip–surface distance $D$ decreases; curves b, c, and d are calculated for three different values of $D$ and $u$, respectively [in the NC-AFM the quantities $D$ and $u$ are closely related, Eq. (17)]. Special attention has to be paid to the NC-AFM where $P$ and $I$ gains are usually adjusted for an infinite tip–surface distance $D$, since approaching the surface may lead to instabilities (Sec. IV).

### B. Shift in frequency and damping signal in the NC-AFM

Assuming there is a couple $(P-I)$ inside the stability domain, it is now possible to write the shift in frequency $\Delta f$ and the damping signal, the two experimental quantities recorded in NC-AFM mode.

Under steady state conditions, the shift in frequency $\Delta f=\omega_0/2\pi(u-1)=f_0(u-1)$ is deduced from Eqs. (12):

$$\Delta f=f_0\left(\frac{2-[\kappa/3(D^2-Z_m^2)^{3/2}]}{(2-[\cot(\theta_s)/Q])}-1\right). \tag{16}$$

Equation (16) is obtained by assuming that $(1-u^2)\approx 2(1-u)$, i.e., $u\approx 1$ in Eqs. (12). This approximation is reasonable since in many cases $\Delta f/f_0$ is much less than $10^{-3}$. The magnitude of the shift in frequency depends on tip–surface interaction and it is weakly dependent on phase $\theta_s$ as long as $\cot(\theta_s)/Q \ll 2$.

In the NC-AFM, the Barkhausen criterion requires that $(\theta_s+\phi)=2n\pi$, where $\phi$ is set with the phase shifter in Fig. 2(a). From a practical point of view, phase $\phi$ exhibits a weak frequency dependence (see a detailed analysis in Sec. IV). Thus, for the particular case of $\theta_s=-\pi/2$, i.e., $\Delta f=0$ for $D\rightarrow\infty$, the shift in frequency $\Delta f$ does not depend on the quality factor $Q$, and $\Delta f$ is given by

$$\Delta f=-f_0\frac{\kappa}{6(D^2-Z_{ms}^2)^{3/2}}. \tag{17}$$

If $\theta_s\neq-\pi/2$, the shift in frequency depends on the damping coefficient $\gamma=1/2Q$, and misinterpretation is then possible because $\Delta f$ is not the sign of the nondissipative term of the interaction, so the case of $\theta_s\neq-\pi/2$ has to be avoided. On the other hand, oscillation at $\theta_s\neq-\pi/2$ would require a high value of $G$ and/or damping signal $D_{\mathrm{amp}}$.

A typical plot of the shift in frequency $\Delta f$ [Eq. (17)] versus the distance $D$ for $\theta_s=-\pi/2$ is shown in Fig. 4(a) (dashed curve a).

From Eq. (17), it appears that the high sensitivity of the NC-AFM is not related to the high value of $Q$ as is sometimes mentioned in the literature. However, the phase noise density of the loop varies as $1/Q^2$,[29] so a high value of $Q$ is required to obtain good resolution in the NC-AFM. At this stage, we have to point out that the sensitivity is also dependent on the bandwidth $B$ of the frequency demodulator; a small value of $B$ reduces the noise but, as usual, slows down the dynamical response of the frequency demodulator.

Under steady state conditions, the damping signal $D_{\mathrm{amp}}$ is deduced from Eqs. (12) as

$$D_{\mathrm{amp}}=-\frac{Z_{ms}u}{Q}\frac{k_c}{F_0}\frac{1}{\sin(\theta_s)}. \tag{18}$$

The closed loop in Fig. 2(a) shows that $F_0=k_1GA_0k_2Z_{ms}$; thus

$$D_{\mathrm{amp}}=-\frac{1}{Q}\frac{uk_c}{k_1k_2GA_0}\frac{1}{\sin(\theta_s)}. \tag{19}$$

For $Q=2\gamma=2(\gamma_0+\gamma_{\mathrm{int}})$, with $\gamma_0$ the damping coefficient when $D\rightarrow\infty$ and $\gamma_{\mathrm{int}}$ the damping coefficient related to tip–surface interaction, respectively, it is clear that the damping signal appears to be a good measure of the dissipative term of tip–surface interaction. However, the damping signal in Eq. (19) is also frequency dependent through the $u$ term. Thus, a change in the nondissipative term of tip–surface interaction leads also to a change in damping signal: $\Delta D_{\mathrm{amp}}/D_{\mathrm{amp}}=\Delta u/u$. In order to avoid any ambiguity about the dissipative term of tip–surface interaction, the damping signal has to be treated simultaneously with the shift in frequency signal.

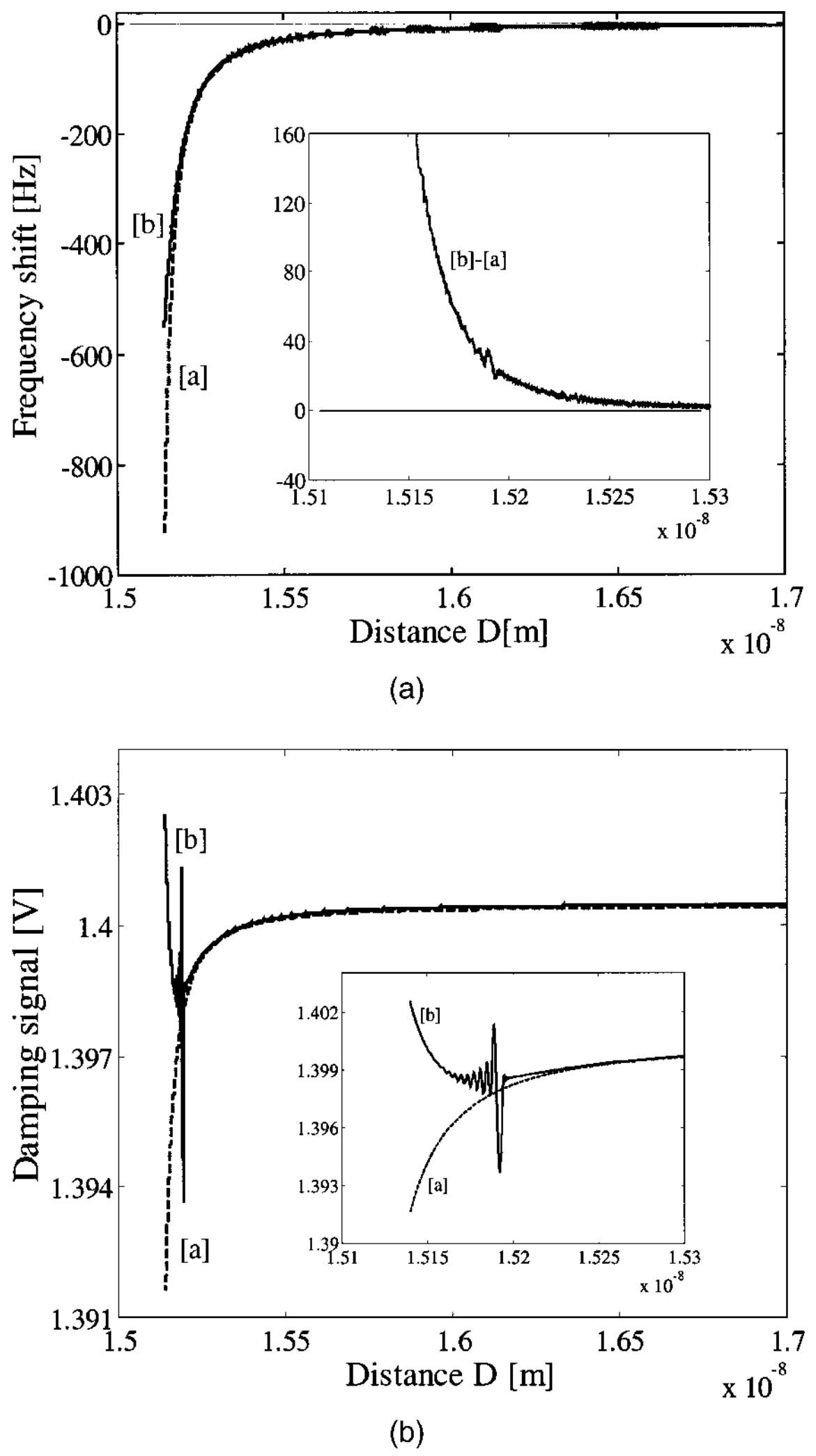

FIG. 4. Shift in frequency $\Delta f$ (a) and damping signal (b) vs tip–surface distance $D$ in the NC-AFM. Curves a in (a) and (b) are given by Eqs. (17) and (19), and curves b are given by the virtual machine. The insets in (a) and (b) are the magnifications around the beginning of the instability of the shift in frequency and the damping signal. It is important to note the correlation between the two signals.

A typical plot of the damping signal [Eqs. (17) and (19)] versus distance $D$ for $\theta_s=-\pi/2$ is given in Fig. 4(b) (dashed curve a).

## IV. COMPARISON WITH THE NC-AFM VIRTUAL MACHINE

Equations (17) and (19) are established by assuming that steady state conditions are fully satisfied; these equations give no information about dynamical solutions. From a practical point of view, the steady state conditions require an infinitely slow sweep rate. If the steady state conditions are not fully satisfied, the $\Delta f$ and $D_{\mathrm{amp}}$ signals depend on proportional $P$ and integral $I$ gains of AGC. The set of equations, Eqs. (8), (10), and (11), has no analytical solution. As an alternative to this problem, a virtual NC-AFM machine helps to evaluate oscillation behavior that cannot be accounted for by the steady state approximation. Both virtual and hardware machines can be represented in block diagram form as shown in Fig. 2(a). The hardware NC-AFM machine is a hybrid machine made of a Digital Instruments head and a controller (Nanoscope E)[30] and Omicron electronics (AFMCU).[31] The commercial Digital Instruments machine was modified for use in noncontact mode. For more details concerning the virtual machine, see Ref. 13.

The virtual machine is in fact a set of differential equations that describe each block in Fig. 2(a). The set of equations is numerically solved without any approximation.

The OTCS is still described by Eq. (1). The peak detector in Omicron electronics is made of a rectifying diode, a $R-C$ circuit, and a first-order low pass filter. For simplicity a quadratic detector and a second-order low pass filter are used in the virtual machine. The output $V(t)$ of the peak detector is given by

$$V(t)=\sqrt{2}[W(t)]^{1/2},$$

with

$$\frac{1}{\omega_c^2}\ddot{W}+\frac{1}{Q_c\omega_c}\dot{W}+W=[k_2A_0z(t)]^2, \tag{20}$$

where $f_c=\omega_c/2\pi$ and $Q_c$ are the center frequency and the quality coefficient of the low pass filter.

Assuming a steady state solution, $z(t)=Z_{ms}\cos(\omega t)$; the output of the peak detector can be written approximately as: $V(t)\approx k_2A_0Z_{ms}$ if $f_c$ satisfies the inequality $f_c\ll\omega/2\pi$. The role of the filter is to partially eliminate the component at $2\omega$. However a compromise has to be found between the magnitude of the component at $2\omega$ and the settling time of the output $V(t)$.

Virtual and hardware AGC are described by Eqs. (9) and (10). The phase shifter ($\phi$) is a second-order all-pass filter and the output $y(t)$ obeys the differential equation,

$$\tau_d\frac{d^2y(t)}{dt^2}+2\frac{dy(t)}{dt}+\frac{1}{\tau_d}y(t)$$
$$=Gk_2A_0\left[\tau_d\frac{d^2[z(t)D_{\mathrm{amp}}(t)]}{dt^2}-2\frac{d[z(t)D_{\mathrm{amp}}(t)]}{dt}\right.$$
$$\left.+\frac{1}{\tau_d}z(t)D_{\mathrm{amp}}(t)\right]. \tag{21}$$

Assuming a steady state solution, $z(t)=Z_{ms}\cos(\omega t)$, and thus a constant damping signal $D_{\mathrm{amp}}$, the general expression for $y(t)$ is

$$y(t)=Y_{ms}\cos(\omega t-\phi), \tag{22}$$

where $Y_{ms}=Gk_2A_0D_{\mathrm{amp}}Z_{ms}$ and $\phi=4\,arctg(\tau_d\omega)$.

Rapid calculation shows that variation of the phase $\Delta\phi$ of the phase shifter is related to the variation in frequency $\Delta f$ according to $\Delta\phi\approx-8\pi\tau_d/1+(\omega_0\tau_d)^2\Delta f$. Assuming, for instance, $\Delta f=1000$ Hz, which is a very large value, then $\Delta\phi\approx-9.4\times10^{-3}$ rad for $\omega_0=9.255\times10^5$ rad s$^{-1}$ and $\tau_d=2.608\times10^{-6}$ s (the values used for the calculations here in Sec. IV). Thus, when the loop is closed like in Fig. 2(a), we consider that phase $\theta$ of the OCTS is kept at an approximately constant value because the Barkhausen criterion requires that $(\theta+\phi)=2n\pi$.

Finally, the feedback driving force $F_{\mathrm{exc}}(t)$, in Fig. 2(a), is given by

$$F_{\mathrm{exc}}(t)=k_1y(t). \tag{23}$$

The set of equations, Eqs. (1), (9), (20), (21), and (23), is numerically solved using the Simulink tool box in Matlab and a Runge–Kutta method, the fixed step size $\Delta t$ used for the calculations being about $2\pi/(70\omega_0)$. To start the oscillation in the closed loop, a very short pulse is applied at the input of the OTCS [not shown in Fig. 2(a)].

### A. Shift in frequency and damping signal with the NC-AFM virtual machine

Curves b in Figs. 4(a) and 4(b) are the shift in frequency and the damping signal versus distance $D$, respectively. The approach rate is slow, about 1.2 nm s$^{-1}$, in order to keep the system under adiabatic conditions. These curves have to be compared to the theoretical curves (a) obtained from Eqs. (17) and (19). The damping signal becomes unstable when distance $D$ is less than ≈15.2 nm [see the inset in Fig. 4(b)]. Parameter $A_0$ was adjusted to obtain $Z_{ms}=15.1$ nm and the $P$ and $I$ gains were set to 60 and 35 000, respectively. The limit of stability calculated with the virtual machine is in good agreement with the theoretical results in Sec. III. From the curves in Fig. 3, the limit of stability is found at about 15.17 nm.

The inset in Fig. 4(a) shows that the shifts in frequency given by the virtual machine and by Eq. (17) are in good agreement as long as the shift in frequency is less than about 200 Hz. For shifts in frequency larger than 200 Hz, i.e., for distance $D$ very close to the amplitude $Z_{ms}$ of the oscillation, the difference between the theoretical predictions (curve a) and the virtual machine (curve b) becomes more and more pronounced. Two reasons can explain the difference between curves b and a.

(1) The smaller the $(D-Z_{ms})$ distance, the larger the shift in frequency, and approaching the surface makes the rate of change of instantaneous frequency of the loop very large. The virtual machine uses a quadrature frequency demodulator with 455 kHz intermediary frequency, the same as that is used by Omicron.[31] Thus the output signal of the frequency demodulator has a rise time, $\tau_r$, that is directly related to the bandwidth $B$ of measurement, $\tau_r=1/2\pi B\approx1$ ms. Consequently, the frequency demodulator cannot follow an instantaneous change of frequency.

(2) The damping instability leads to an abrupt change in frequency. This can be understood with the help of curves b, c, and d and the horizontal line Δ in Fig. 1(b). If the damping signal becomes unstable, the amplitude of the oscillations is no longer constant and, because phase $\theta$ is kept constant, the frequency changes, in agreement with Fig. 1(b), where the horizontal line Δ is the locus of the quiescent point for the OTCS. Qualitatively, one can interpret the concomitant change in frequency as follows: while the phase remains constant, as indicated by the Δ line, the AGC loop is unable to keep the oscillation amplitude constant. Because of this, the oscillation reaches another state that corresponds to a transient value of the oscillation amplitude and, consequently, another value of the shift in resonance frequency. The fre-quency instabilities are unambiguously observed close to 15.2 nm [inset in Fig. 4(a)]. There is a strong correlation between the instabilities of the damping signal and those of the frequency. In conclusion, without fully rejecting point (1), we suggest that the main reason for the difference between curves a and b is the decrease in amplitude of the oscillations which is related to instability of the system. The correlation between the damping and the frequency shift instabilities has already been observed in NC-AFM experiments.[32]

The virtual machine is now used to investigate the case where the adiabatic conditions are not satisfied. As mentioned in Sec. III, the shift in frequency and the damping signal now become dependent on the $P$ and $I$ gain. Two stimuli are successively applied to the cantilever: (i) a nondissipative force step and (ii) a dissipative force step. For each step, two pairs of $P$ and $I$ gains are chosen, denoted $P_1$ and $P_2$ in Fig. 3. These two points are inside the stability domain. We want to learn more about the behavior of the machine when small, but fast, perturbation is applied. The distance and the set point amplitude are $D=15.4\,\mathrm{nm}$ and $Z_{ms}=15.1\,\mathrm{nm}$, respectively, thus the stability domain is still very close to the hatched zone in Fig. 3.

### B. Transient responses to a nondissipative force step

A Heaviside function that describes a variation in the step of the Hamaker constant $H$ is applied, with the step magnitude $\Delta H$ 1% of the initial value of $H$. For such a step, the frequency of the loop instantaneously changes. Curves a and b in Fig. 5(a) are the shifts in frequency for two pairs of values $(P\!-\!I)$, respectively. As expected the gains in $P$ and $I$ have no effect on the variation of the shift in frequency. The output of the demodulator is only determined by its bandwidth, thus it exhibits a delay response for time $\tau_r \approx 1$ ms. Curve c gives the instantaneous response [Eq. (17)] that corresponds to an "ideal" machine with $\tau_r=0$.

Curves a and b in Fig. 5(b) are the damping signal for the pairs $P_1$ and $P_2$, respectively; curve c is again the response of an ideal machine [Eq. (19)]. A high value of $P$ and of $I$, like in curve b, leads to weakly damped oscillatory behavior of the damping signal whereas a single overshoot is observed for small values of $P$ and $I$, so $P_1$ is more suitable in the case of a nondissipative force step. However, it is worth noting that the overshoot, although being spread over quite a large time scale (a few ms), remains negligible, about $10^{-4}$ of the initial damping value, which corresponds to additional dissipated energy of $1.9\times10^{-22}\,\mathrm{J}$ $(10^{-4}\,k_c Z_{ms}^2/Q$ with $k_c=40\,\mathrm{N\,m^{-1}}$, $Z_{ms}=15.1\,\mathrm{nm}$, and $Q=4750)$. Such variation cannot be observed except if the experiment is performed at temperature lower than about 10 K $(kT=1.38\times10^{-22}\,\mathrm{J}$ at 10 K).

### C. Transient responses to a dissipative force step

The dissipative force step is obtained by using a Heaviside function for the $\gamma$ coefficient. Curves a and b in Fig. 6(a) are the damping signals for the pairs $(P\!-\!I)$, respectively. Curve c is for an ideal machine, given by Eq. (19). The transient response of the damping signal has the same features as those above; high values of $P$ and $I$ lead to weakly damped oscillatory behavior (curve b) whereas a single overshoot is observed for small values of $P$ and $I$ (curve a).

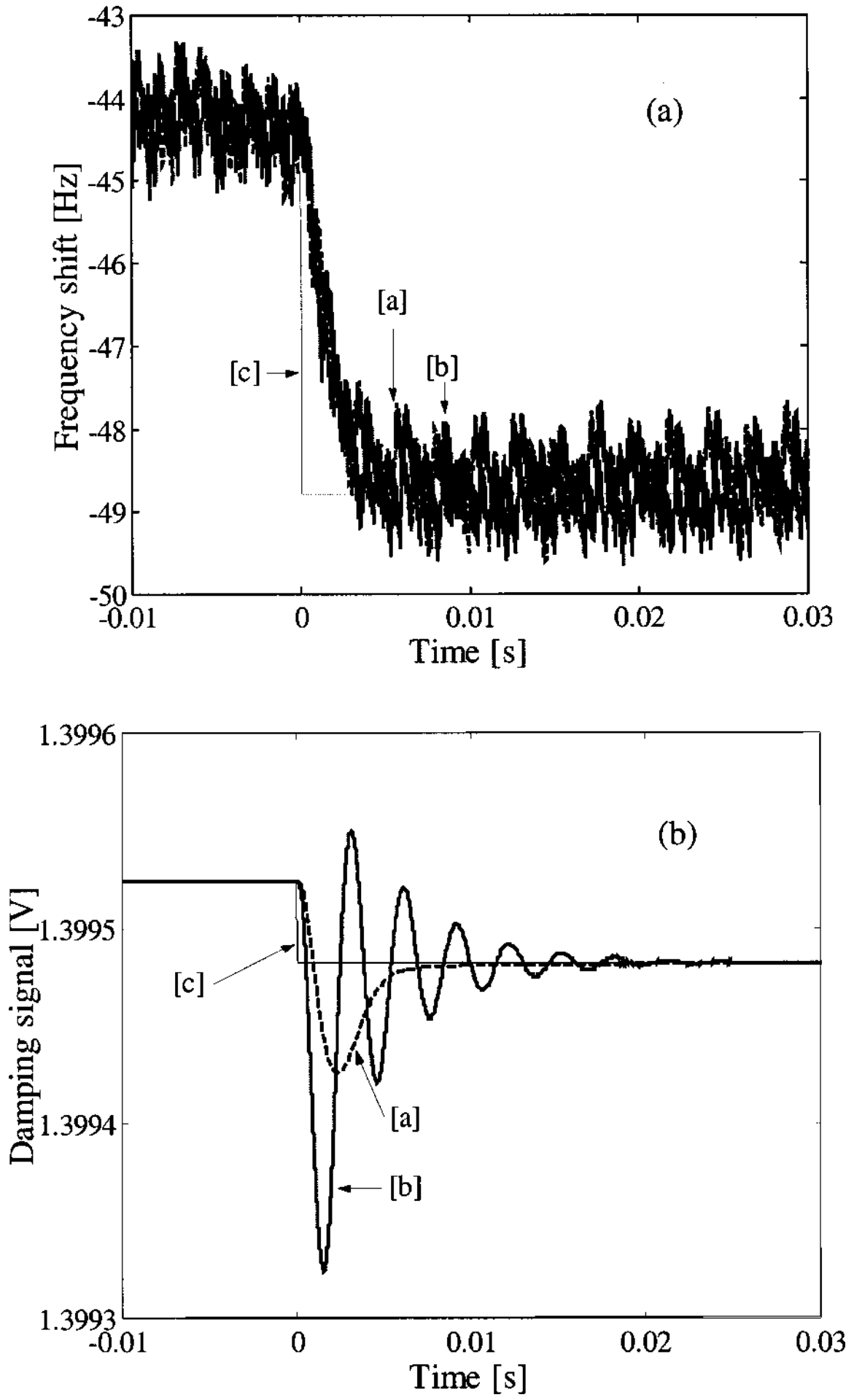

FIG. 5. Transient responses of the shift in frequency and damping signal for a nondissipative force step. Curves a and b are obtained for $P=20$, $I=1000$ and $P=60$, $I=35\,000$, respectively; curves c are given by Eqs. (17) and (19).

Under steady state conditions, the shift in frequency cannot be related to the dissipation term [curve c in Fig. 6(a)]. The shift in frequency remains constant [Eq. (17)]. Curves a and b are the shifts in frequency given by the virtual machine. Small values of $P$ and $I$, like in curve a, lead to large variation of the frequency. This behavior is still consistent with the explanations provided in (2) in Sec. IV A, where the change in frequency is induced by the amplitude of the oscillations not being kept constant because AGC is not able to correct amplitude fluctuations quickly enough. Therefore, the $P_2$ would be more appropriate in the case of a dissipative force step to avoid misinterpretation about the shift in frequency of the signal.

These two examples show that it is not very easy to separate contributions of dissipative and nondissipative force if adiabatic conditions are not satisfied; there is no ideal value for the gains in $P$ and $I$. However, these two examples can be used as a guide to avoid misinterpretation when using the NC-AFM.

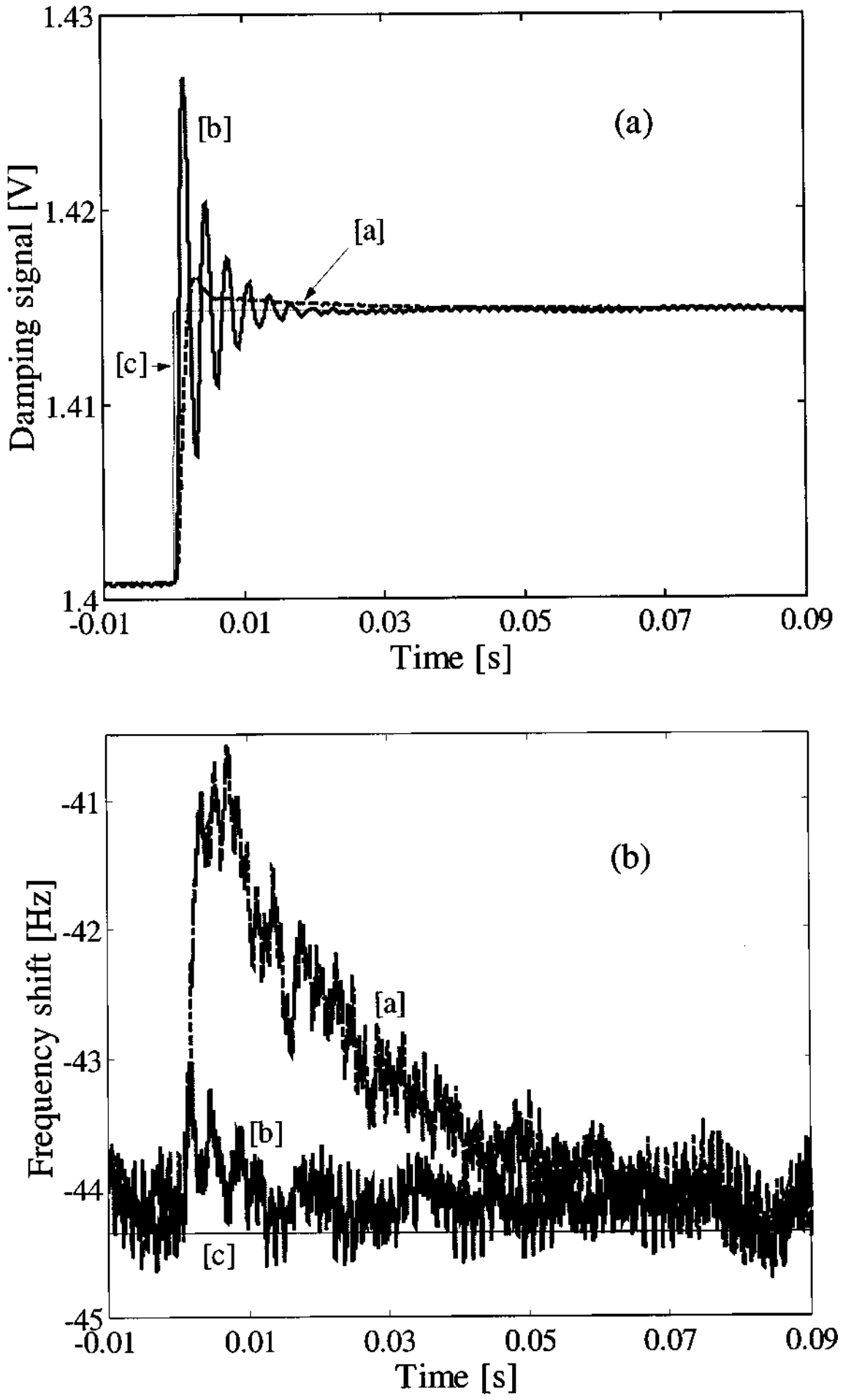

FIG. 6. Transient responses of the shift in frequency and damping signal for a dissipative force step. Curves a and b are obtained for $P=20$, $I=1000$ and $P=60$, $I=35\,000$, respectively; curves c are given by Eqs. (17) and (19).

## V. DISCUSSION

By solving the equations of motion of the OTCS and those of CAG and using the Routh–Hurwitz criterion, we showed that the NC-AFM is stable if proportional $P$ and integral $I$ gains are inside the stability domain. The stability domain is strongly dependent on tip–surface interaction. Only pure van der Waals attractive interaction was considered here, but the method is valid for any kind of interaction.

Also deduced from the equations of motion of the OTCS were the shift in frequency and the so-called damping signal under steady state conditions. It was shown that the shift in frequency is related only to nondissipative force if the phase of the OTCS is $-\pi/2$. The damping signal depends on losses of the cantilever but it is also dependent on the frequency of the loop, so special attention has to be paid to interpreting this signal to avoid misinterpretation.

To validate theoretical results, which were obtained using some minor approximations, a virtual NC-AFM machine was built. The virtual machine, which is identical to a hardware machine, is in fact a set of differential equations solved numerically without any approximating. The shift in frequency and damping signal given by the virtual machine in approach–retract mode are in good agreement with the theoretical results as long as conditions of stability are satisfied. The stability domain given by the virtual machine and theoretical results are in good agreement. We have also explained the correlation between the oscillations of the damping signal and those of the shift in frequency.

In scanning or approach–retract mode, on which adiabatic conditions are seldom satisfied, it is of primary importance to characterize the dynamical behavior of the NC-AFM, which depends on proportional $P$ and integral $I$ gains. Starting from the equations of motion of the OTCS and the equations of CAG, it is rather difficult to obtain information about the dynamical behavior. An alternative to this problem is use of the virtual machine which is a very powerful tool with which to study, for instance, transient responses of the shift in frequency and the damping signal. We have shown that a step of nondissipative force gives an unexpected change in the damping signal; conversely, a step of dissipative force gives an unexpected change in frequency. In both cases, the transient responses are completely influenced by the choice of proportional $P$ and integral $I$ gains of CAG. There is no ideal value for the pair $(P-I)$, so care has to be taken in interpreting transient responses of the shift in frequency and damping signal.